# Bio-realistic and versatile artificial dendrites made of anti-ambipolar transistors

*Yifei Yang†, Mingkun Xu†, Jing Pei, Peng Li, Guoqi Li, Si Wu, Huanglong Li\**

Y. Yang, M. Xu, J. Pei, H. Li
Department of Precision Instrument, Center for Brain Inspired Computing Research, Tsinghua University, Beijing, 100084, China.
E-mail: li_huanglong@mail.tsinghua.edu.cn

P. Li
State Key Laboratory of Precision Measurement Technology and Instruments, Department of Precision Instrument, Tsinghua University, Beijing 100084, China.

G. Li
Institute of Automation，Chinese Academy of Sciences; Beijing, 100190, China.
University of Chinese Academy of Sciences, Beijing, 100049, China.

S. Wu
School of Psychology and Cognitive Science, IDG/McGovern Institute for Brain Research, Center for Quantitative Biology, PKU-Tsinghua Center for Life Sciences, Peking University, Beijing, 100871, China.

H. Li
Chinese Institute for Brain Research; Beijing, 102206, China.

**Abstract**
The understanding of neural networks as neuron-synapse binaries has been the foundation of neuroscience, and therefore, the emerging neuromorphic computing technology that takes inspiration from the brain. This dogma, however, has been increasingly challenged by recent neuroscience research in which the downplayed dendrites were found to be active, dynamically unique and computationally powerful. To date, research on artificial dendrites is scarce and the few existing devices are still far from versatile or (and) bio-realistic. A breakthrough is hampered by the limited available physical mechanisms in mainstream device architectures. Here, we experimentally demonstrate a bio-realistic and versatile artificial dendrite made of WSe$_2$/MoS$_2$ heterojunction-channel anti-ambipolar transistor, which can closely mimic the experimentally recorded non-monotonic dendritic Ca$^{2+}$ action potential that underpins various sophisticated computations. Spiking neural network simulations reveal that the incorporation of this nonconventional but bio-realistic dendritic activation enhances the robustness and representational capability of the network in non-stationary environments. By further exploiting the memristive effect, dendritic anti-ambipolar transistor can naturally mimic $Ca^{2+}$-mediated regulation of synaptic plasticity (meta-plasticity) at dendritic spine, an important homeostatic phenomenon due to dendritic modulation. The invention of dendritic anti-ambipolar transistor completes the family of neuromorphic transistors and represents a major advance in diversifying the functionality of transistors.

## 1. Introduction

The past half century has witnessed the largest man-made growth rate of any kind in the ~10,000 years of human civilization[1], that is, the exponential growth of the speed of digital computer. Much of this performance improvement comes from the miniaturization of the building-block computer component, i.e., transistor. After the year 2000, however, transistor miniaturization has slowed down because its physical limits have been being approached[2]. In this context, incorporation into transistors of functionalities that provide additional value in different ways has become more widely pursued.

Although digital computer is in part the product of human's pursuit for building the brain, it differs from the brain significantly. In fact, digital computers, though faster at doing logical computations, have traditionally lagged behind the brain in key areas, such as pattern recognition, adaptivity and fault tolerance. To meet the rising demand for machine intelligence with higher flexibility and efficiency, the brain neural network has become an important source of inspiration for the development of next-generation computing technology, i.e., neuromorphic computer. In addition to the apparent differences in computing architecture and algorithm, a neuromorphic computer also takes advantage of nonconventional devices that can physically implement similar functionalities of the brain components to accelerate brain-inspired algorithms. This gives rise to a new class of transistors known as the neuromorphic transistors [3-6].

Based on the prevailing understanding of neural networks as neuron-synapse binaries (bi-NNs) since the pioneer work of neurophysiologist W. McCulloch and mathematician W. Pitts in 1943,[7] main efforts in developing neuromorphic transistors have been directed towards synaptic transistors[8-20] and neuro-transistors[21-23]. Interestingly, both types of neuromorphic transistors share similar underlying physical mechanisms: for synaptic transistors, memristive mechanisms are engineered into the devices to realize gate input (as pre-synaptic input) history-dependent channel conductance (under zero gate voltage) modulations, therefore emulating the long-term plasticity of biological synapses; while for neuro-transistors, parallels are often drawn between the memristive processes of channel conductance evolutions and the processes of neuronal integrations and activations. The channel conductances can be either capacitively controlled via metal-oxide-semiconductor (MOS) capacitors as charge distributions in gate dielectrics have changed, or atomically controlled as the atomic structures or atomic compositions of the channels themselves have changed. The induction of synaptic plasticity (either potentiation or depression) and monotonic neuronal activation (mathematically modelled by a step function or sigmoid or ReLU function) requires monotonic dependence of the channel conductance on the accumulated gate inputs. This requirement can be naturally satisfied in typical unipolar (or homojunction-channel) transistors as long as charge redistributions in MOS capacitors (or atomic structure changes in the channels) are unidirectional in response to individual gate inputs of given voltage polarities, as they typically do.

This eighty-year old neuroscience dogma (neural networks as neuron-synapse binaries), however, has been increasingly challenged by recent neuroscience research. In fact, when looking at any neuroscience textbook illustration, one will notice that the neuron cell bodies are often depicted to have many branched extensions, resembling trees. These are dendrites, discovered more than a century ago[24-27]. Traditionally, dendrites have been viewed as nothing but simple passive cables. Over the past few decades, however, mounting research evidences have been accumulated, pointing to the fact that dendrites shoulder more considerable computational responsibilities than once imaginable[24-27]. Therefore, it has become increasingly obvious that dendrites can in no way be downplayed and they are an important source of inspiration for empowering artificial neural networks and neuromorphic computing.

Dendritic computation has generally been categorized into two types: passive and active. Passive dendritic computation performs summation and filtering of the synaptic input signals, linearly or nonlinearly[24-27]. Richly endowed with voltage-dependent ion channels, dendrites can also compute actively. One of the most important biochemical bases of active dendritic computation is the generation of dendritic $Ca^{2+}$ action potentials (dCaAPs)[24-27]. Unlike neuronal action potential that is mainly generated via rapid $Na^+$ influx, dCaAPs have longer durations and can deliver substantial amount of charges to the neuron cell body, thereby influencing the generation of neuronal action potentials.

The active properties of dendrites, in particular, dCaAP generation, allow them to perform important logical operations, such as AND, OR, and AND-NOT, which underlie coincidence detector function, feature selectivity, and so on[24-27]. Very recently, dCaAP in layer 2/3 pyramidal neurons of the human cerebral cortex ex vivo has been found to have non-monotonic amplitude with respect to the intensity of input stimulus[28]. This enables the dendrite unit to solve the even more challenging exclusive-OR (XOR) problem, a computation that was considered solvable only by networks of neurons[29].

To date, research on artificial dendrites is scarce and the few existing devices are still far from versatile or (and) bio-realistic. In early dendritic transistors[30-33] and recent two-terminal dendritic devices[34,35], only passive dendritic computational functionalities, such as filtering and supralinear/sublinear summation, have been demonstrated. These emulated cellular-level behaviors, however, still lack sub-cellular-level bio-explainability, which may therefore prevent device functionalities from being extended to the level of elaboration and sophistication as their biological counterparts, as recently implied by our prototype demonstrations of bio-explainable synaptic devices[11] and neuro-devices[36]. As the harbors for synapses, dendrites are also the sites where synaptic plasticity, the cellular foundation of learning and memory, occurs. However, demonstrations of plasticity functions as controlled by dendritic mechanisms are lacking.

Though these limitations are obvious, a breakthrough in artificial dendrite device technology is still hampered by the limited available physical mechanisms in mainstream device architectures. For example, two-terminal memristors are passive without gain and therefore cannot be used for emulating active dendritic computational functions; while for classic transistors based on homojunction semiconducting channels, their unipolar electron or hole transport mechanism results in monotonic dependence of the channel conductances, transient (upon the arrival of gate inputs) or steady (under zero gate biases), on the accumulated gate inputs, which is not suitable for emulating non-monotonic dendritic activation.

Motivated by the quest for the dendritic function in neuromorphic platforms, we here propose a generic, compact and CMOS-compatible device architecture for dendritic function implementation, that is, anti-ambipolar (AAB) transistor with p-n heterojunction channel[37,38]. The novel carrier transport phenomenon in dendritic AAB-transistor can be used to closely mimic the experimentally recorded non-monotonic dCaAP. Spiking neural network (SNN) simulations reveal that the incorporation of this nonconventional but bio-realistic dendritic activation enhances the robustness and representational capability of the network in non-stationary environments compared to conventional SNN. By further exploiting the memristive effect, dendritic AAB-transistor can naturally mimic $Ca^{2+}$-mediated regulation of synaptic plasticity (meta-plasticity) at dendritic spine, an important homeostatic phenomenon due to dendritic modulation. The invention of dendritic AAB-transistor completes the family of neuromorphic transistors and represents a major advance in diversifying the functionality of transistors.

## 2. Results and discussion
### 2.1. The emulation of non-monotonic dCaAP in dendritic AAB-transistor

The proposal is based on our identification of a natural parallel between the AAB behavior and the dCaAP phenomenon, as illustrated in **figure 1a,b**. A back-gated n-type $MoS_2$/p-type $WSe_2$ heterojunction-channel transistor is fabricated by the photolithography, magnetron sputtering and lift-off processes on the $SiO_2/p^{++}$-Si substrate (see Methods). **Figure 1c** shows the optical image and the atomic force microscopy (AFM) image of the heterojunction area. Two control devices, namely, $MoS_2$ and $WSe_2$ homojunction-channel transistors, are also fabricated.

Raman spectroscopies are used to characterize the properties of the deposited $MoS_2$ and $WSe_2$ films. For the $MoS_2$ film, as shown in **figure 1d**, two characteristic peaks at ~ 380 $cm^{-1}$ and ~ 405 $cm^{-1}$ are observed, in consistence with those of the in-plane Mo-S vibrational mode ($E^1_{2g}$) and the out-of-plane S-S vibrational mode ($A_{1g}$) in stacked crystalline $MoS_2$, respectively[39]. To further confirm the chemical states of the deposited films, X-ray photoelectron spectroscopy (XPS) analyses are also preformed. **Figure 1e, f** display the spectra corresponding to the Mo 3d band and the S 2p band for the $MoS_2$ film, where the characteristic Mo-S binding energies are identified[40], including the 233.5 eV for Mo $3d_{3/2}$, 230.7 eV for Mo $3d_{5/2}$, 163.2 eV for S $2p_{1/2}$, and 162.1 eV for S $2p_{3/2}$. For the WSe$_2$ film, as shown in **figure 1g**, a broad peak centered at ~ 246 cm$^{-1}$ is observed in Raman spectrum, in consistence with the two overlapped ones corresponding to the in-plane W-Se vibrational mode ($E^1_{2g}$) and the out-of-plane Se-Se vibrational mode ($A_{1g}$) in stacked crystalline WSe$_2$[41,42], respectively. **Figure 1h** displays featured XPS peaks at 31.6 eV and 34.4 eV attributed to the doublet of W $4f_{7/2}$ and W $4f_{5/2}$, respectively. In **figure 1i**, the doublet peaks of Se $3d_{5/2}$ and Se $3d_{3/2}$ at 52.8 eV and 54.9 eV, respectively, can also be deconvoluted[42].

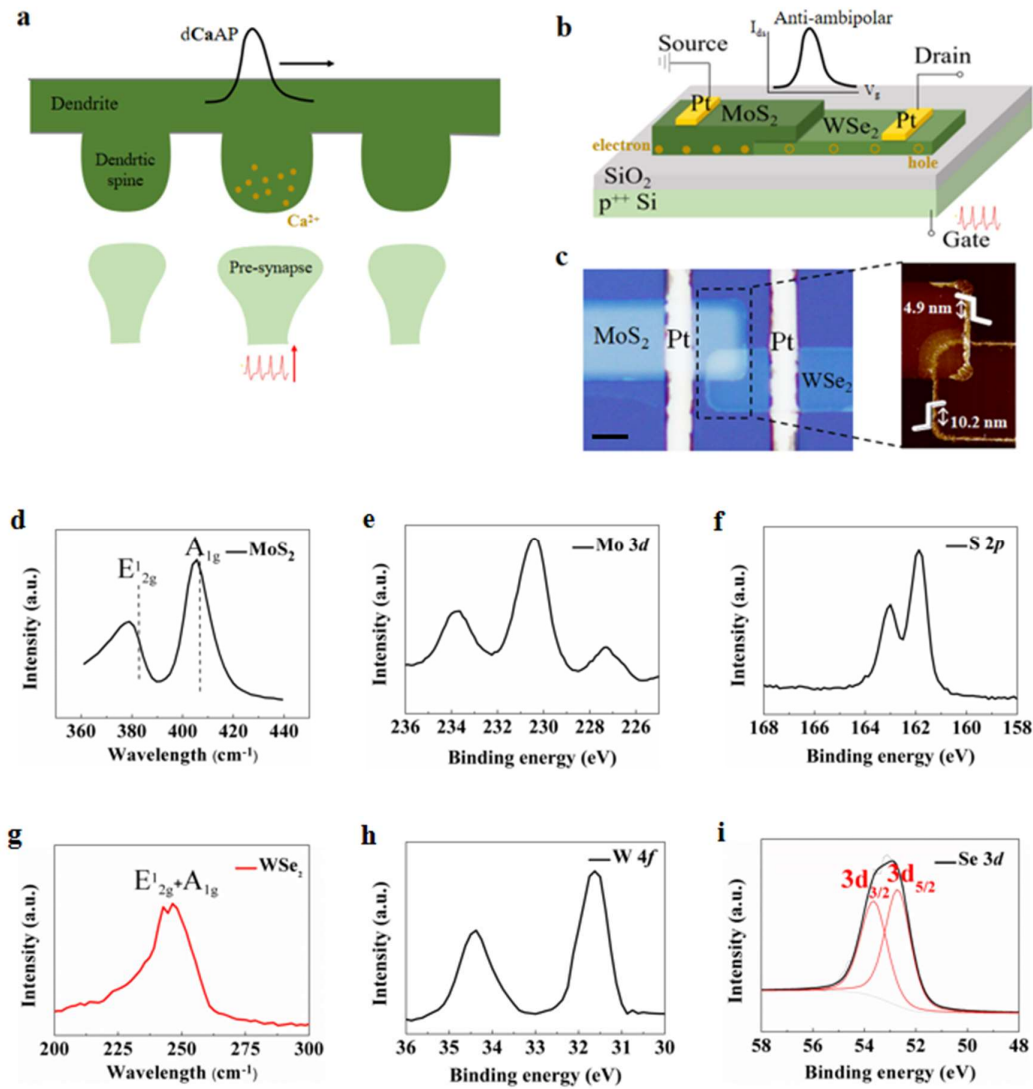

**Figure 1. Device architecture and materials characterizations. a.** Schematic of a biological dendrite with the ability to generate dCaAP. **b.** Device schematic of the dendritic AAB-transistor with MoS$_2$/WSe$_2$ heterojunction-channel. **c.** Optical image of the AAB-transistor (scale bar: 5 μm) and the corresponding AFM image of the heterojunction area. **d.** Raman spectrum of the MoS$_2$ part of the heterojunction channel. **e.** Mo $3d$ XPS and **f.** S $2p$ XPS spectra of the MoS$_2$ part of the heterojunction channel. **g.** Raman spectrum of the WSe$_2$ part of the heterojunction. **h.** W $4f$ XPS

and **i.** Se 3*d* XPS spectra of the WSe$_2$ part of the heterojunction channel.

Electrical measurements are performed to obtain the transfer characteristics of the heterojunction-channel transistor and the two control homojunction-channel transistors (see Methods). Supplementary figure S1 shows the transfer curves of the MoS$_2$ and WSe$_2$ homojunction-channel transistors, where typical unipolar n-type and p-type conductivities, respectively, are clearly seen. **Figure 2a** shows the transfer curves of the WSe$_2$/MoS$_2$ heterojunction-channel transistor under positive (forward bias) and negative (reverse bias) supply voltages ($V_{ds}$). Under negative $V_{ds}$, the transistor cannot be turned on regardless of the applied $V_g$. Only under positive $V_{ds}$ can the transistor show nontrivial transfer characteristic. To be specific, the transistor is in the off state when $V_g$ is positive and large, while it gradually turns on as $V_g$ decreases till $V_g$ polarity has been reversed. The source-drain current ($I_{ds}$) peaks at $V_g \sim$ -4.8 V but decreases if $V_g$ becomes more negative. Overall, $I_{ds}$ is non-monotonically dependent on $V_g$, resulting in the so-called AAB transfer curve.

The AAB behavior can be understood in a fairly plain way that the p-n heterojunction channel is effectively the tandem of a p-type channel and an n-type channel, which are commonly gate-biased, resembling the structure of a CMOS inverter (**figure 2b**). In our case, $I_{ds}$ under large positive (negative) $V_g$ is limited by the high resistance of the p-WSe$_2$ (n-MoS$_2$) channel despite that the n-MoS$_2$ (p-WSe$_2$) channel is sufficiently conductive. Only at an appropriate $V_g$ can both the p-WSe$_2$ and n-MoS$_2$ parts of the heterojunction channel be reasonably conductive, giving rise to $I_{ds}$ peak. The output curves also correspond with the AAB feature, as shown in **figure 2c**, that $I_{ds}$ increases most rapidly under $V_{ds}$ sweep from negative to positive polarity when $V_g$ is -5 V, but less with increasing deviation of $V_g$ from -5 V.

The transfer curve obtained by quasi-DC measurements can only reflect instantaneous dendritic response. To demonstrate dendritic temporal integration, AC pulse train measurements are conducted to investigate the time evolution of $I_{ds}$ output. **Figure 2d** shows the $I_{ds}$ evolution under successive triangular negative $V_g$ pulses of the width of 100 ns, the interval of 100 ns and the intensity of 10 V. In the beginning, the $I_{ds}$ response to $V_g$ pulse gradually increases over time. As just described, $I_{ds}$ of our pristine AAB-transistor peaks at negative $V_g$ region (figure 2d), indicating that the conduction in the absence of $V_g$ is mainly limited by the p-part (WSe$_2$) of the channel. In pulse train measurements, short negative $V_g$ pulses enhance the conductivity of the p-part channel by pumping out holes, but at the expense of some degradation of the conductivity of the n-part (MoS$_2$) by expelling electrons. The gradual increase in $I_{ds}$ with each successive pulses is mainly due to hole accumulation in WSe$_2$ because the pumped out holes cannot completely relax back in the short pulse intervals. However, the increase in $I_{ds}$ does not last long before it reaches the peak value from which the decrease takes place. The decrease can be understood as arising from the continually decreased conductivity of the n-part channel that begins to limit $I_{ds}$ even if the p-part channel has become sufficiently conductive. The decrease continues as more and more electrons are expelled due to incomplete relaxation during the pulse intervals. Therefore, the evolution of $I_{ds}$ is non-monotonically dependent on the number of input $V_g$ pulses, resembling dCaAP. Intriguingly, various forms of activation functions can be emulated under similar pulse train stimulus that only differ in pulse intensity (see supplementary note 1 and supplementary figure S2).

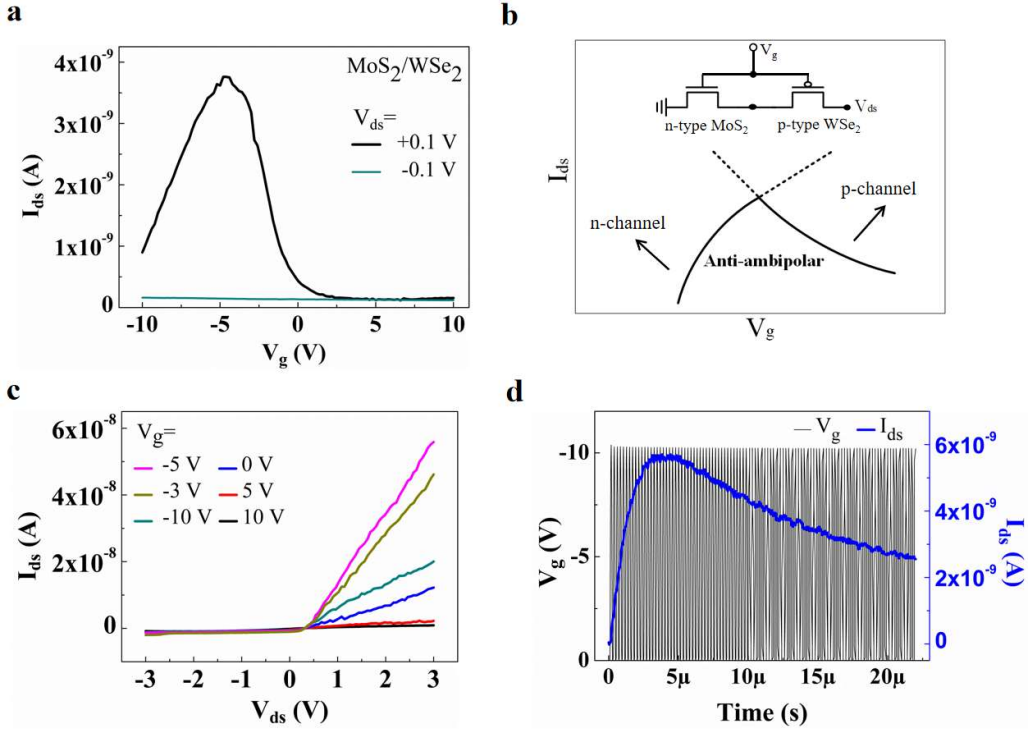

**Figure 2. Electrical measurements of the device. a.** Transfer characteristics of the $MoS_2/WSe_2$ heterojunction-channel transistor under different $V_{ds}$ polarities. **b.** Schematic diagram of the origin of the AAB transfer characteristic. **c.** Output curve of the $MoS_2/WSe_2$ heterojunction-channel transistor under different $V_g$. **d.** Pulse train measurement of the $WSe_2/MoS_2$ heterojunction-channel transistor under the pulse amplitudes of -10 V.

## 2.2. SNN as a synapse-dendrite-neuron trinity with neuro-inspired non-monotonic dendritic activation function

To demonstrate the computational advantages of non-monotonic dendritic activation functions in bio-realistic SNNs, we simulate three SNN models, i.e., conventional SNN as a synapse-neuron binary (bi-SNN) and SNNs as synapse-dendrite-neuron trinities with dendritic activation functions (dAF) simulated by monotonic ReLU function (tri-SNN(R)) and non-monotonic Gaussian function (tri-SNN(G)), respectively, and compare their performance differences. The synapse-dendrite-neuron trinity unit as the building block of our tri-SNN(G) model is schematically shown in **figure 3a**. The dendritic extension is functionalized to be able to integrate synaptic inputs $s_{in}$, spatially and temporally, into its instantaneous membrane potential $V_d(t)$ with certain degree of leakage over time. Dendritic output or activation $V_{dAF}$ is simulated by a non-monotonic Gaussian function of $V_d$. This process mimics the newly found dCaAP[28]. Immediately following dendritic processing, leaky integrate-and-fire (LIF) neural processing begins, by integrating dendritic output $V_{dAF}$ into the instantaneous neuronal membrane potential

$V_n(t)$. When $V_n(t)$ exceeds a certain threshold $V_{th}$, neuron emits a spike $s_{out}$ (activated) and its membrane potential is spontaneously reset to a resting-state value $V_{n\_rest}$, ready to be activated again. Detailed mathematical abstraction of the trinity unit can be found in supplementary note 2.

A three-layer tri-SNN(G) model is schematically shown in **figure 3b**. As suggested[43], the addition of dendritic function allows a single neuron to act as a two-layer ANN, with dendrite serving as the hidden processing unit and the neuron cell body as the output unit. Therefore, compared to conventional bi-SNN of similar architecture, tri-SNN(G) is virtually deeper. The basic performance of the three-layer tri-SNN(G) model is evaluated from four independent trials of training on Fashion-MNIST dataset[44] (see Methods), as shown in **figure 3c**. The network is trained with spatio-temporal back-propagation algorithm (STBP) via gradient substitution[45]. The baseline bi-SNN for comparison has the same network architecture and hyper-parameters (not fine-tuned or trained) for non-dendrite (i.e., neuron) units. It can be seen that bi-SNN exhibits large variation from trial to trial, in terms of both the speed of convergence and the level of test accuracy. By contrast, tri-SNN(G) is highly stable that in all four trials it consistently reaches a high accuracy level, outperforming bi-SNN model at its best. Similar tests are also carried out for tri-SNN(R). Two important observations can be noted: first, the addition of the dendrite unit, regardless of whether its dAF is non-monotonic or not, mitigates the performance variation of the network; second, tri-SNN(G) reaches higher level of test accuracy than does tri-SNN(R), indicating that the non-monotonicity of dAF does matter.

To understand these differences among the three SNN models, we conduct an empirical analysis of their loss landscape structures. **Figure 3d,e** compare the trained models in terms of Hessian eigenvalue (HE)[46] and mode connectivity (MC)[47] that capture the local and global aspects of the landscape, respectively[48]. It can be seen that tri-SNN(G) (bi-SNN) has the largest (smallest) HE value, indicating that its loss landscape is the most locally sharp (flat). If one only looked at local sharpness, one could wrongly predict that tri-SNN(G) (bi-SNN) is the poorest (best) performing model. However, when their global properties measured by MC are compared, it is expected, as proven, that tri-SNN(G) outperforms the other two models because of its nearly zero MC value and therefore globally well-connected loss landscape. In particular, the MC value increases in the order of tri-SNN(G), tri-SNN(R) and bi-SNN, while the order is reversed for the HE value. The performance trend shown here is in line with the observation of Yang et al.[48] that for globally poorly-connected loss landscape (here for bi-SNN), the test accuracy can be low even if the model is converged to a locally flat region. Loss landscapes visualization[49] and t-distributed stochastic neighbor embedding (t-SNE) visualization[50] that give more intuitive impression of the representational capabilities of the network models are presented in supplementary figure S3 and discussed in supplementary note 3.

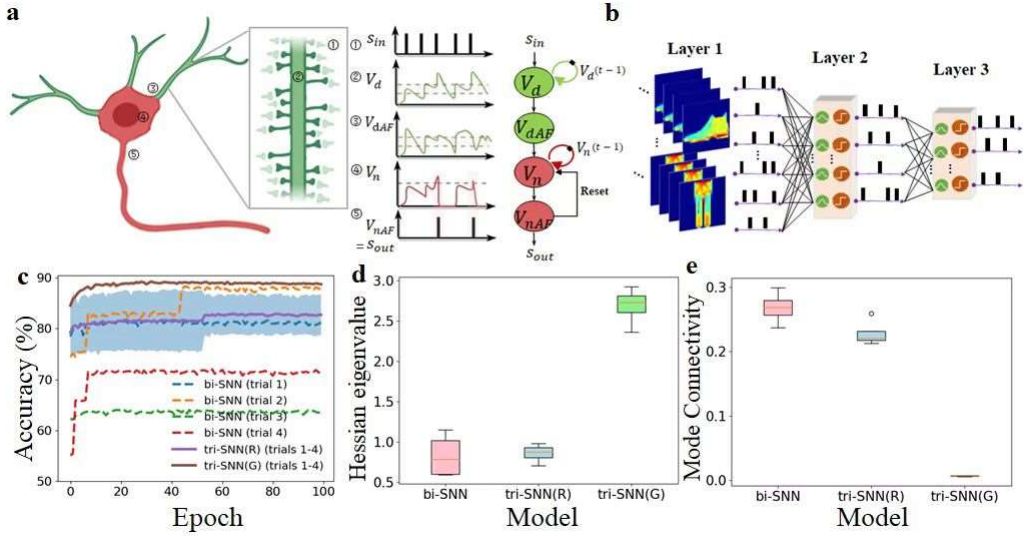

**Figure 3. Comparison of the in-distribution representational performance among bi-SNN, tri-SNN(R) and tri-SNN(G). a**. Schematic illustration of the synapse-dendrite-neuron trinity unit and its mathematical abstraction. **b**. Schematic illustration of the three-layer tri-SNN(G) model. **c**. Test accuracies for bi-SNN, tri-SNN(R) and tri-SNN(G) as functions of epochs. The dark (light) purple and dark brown curves (shadows) are mean accuracies (standard deviations) over four trials for tri-SNN(R) and tri-SNN(G), respectively. **d**. Local properties of the loss landscapes of bi-SNN, tri-SNN(R) and tri-SNN(G). **e**. Global properties of the loss landscapes of bi-SNN, tri-SNN(R) and tri-SNN(G)

A common benchmark for human-like performance in artificial intelligence is the ability to perform robustly in noise-perturbed environments and the continual learning ability to flexibly adapt to changing tasks in a non-stationary world. To benchmark the robustness of tri-SNN to noise perturbations, we train the model on intact Neuromorphic MNIST dataset[51] (see Methods) and test it using salt & pepper noise perturbed test samples. **Figure 4a** shows that the noise-robustness increases in the order of bi-SNN, tri-SNN(R) and tri-SNN(G) over the entire tested noise range. To gain more insight into their robustness properties, we calculate their Cosine distances (**figure 4b**) with respect to the average layer-2 neuron membrane potentials between tests on noise corrupted samples and tests on intact ones. It is seen that the degree of similarity in neuron membrane potential between these two test sets increases in the order of bi-SNN, tri-SNN(R) and tri-SNN(G). The noise resilience of tri-SNN(G) can be associated with the band-pass filter-like Gaussian dAF. In particular, it filters out noises perturbing neuron membrane potential in either direction that tends to make neuron overactive or underactive. Analyses of the neuronal firing patterns in these three SNNs receiving inputs contaminated by noise of different intensities (see supplementary figure S4 and supplementary note 4) give the hint that the computational properties of SNNs are associated with their code transmission properties. Further studies are still necessary to bridge the gap between activity dynamics and network functions.

Currently, most neural network models are trained in narrow and closed task domains

(in-distribution) to exhibit stereotyped input-output mappings. As open world problems where things keep changing over time have become increasingly relevant, neural network models with sufficient flexibility are demanded. However, conventional models generally underperform when presented with changing or incremental data regimes and suffer from rapid performance degradation on earlier leant tasks, known as catastrophic forgetting. **Figure 4c** compares the fifty-task continual learning performance of bi-SNN, tri-SNN(R) and tri-SNN(G) models on standard shuffled MNIST dataset[52] (see Methods). It is clear that tri-SNN(G) consistently outperforms the other two models as the number of learnt tasks exceeds 15, with simple bi-SNN being the poorest performed model. We visualize the patterns of the learnt neural representations for layer-2 neurons by k-means clustering of their membrane potentials[53], as shown in **figure 4d-f**. It is seen that for bi-SNN no easily discernible clustering of neurons occurs; in other words, neural resources are shared across different tasks. In addition, the overall neural activity in bi-SNN is quite low that the membrane potentials of most neurons are close to their resting-state values (zero). In contrast, for both tri-SNN(G) and tri-SNN(R) the layer-2 neurons are, to a certain extent, self-organized into distinct clusters through continual learning, indicating that neurons are task-specific. The reduced representational overlap is beneficial for mitigating the catastrophic forgetting problem. In both tri-SNN models, the non-clustered neurons are in the majority as the background for each task. However, tri-SNN(G) and tri-SNN(R) differ in terms of the relative level of activity of the clustered neurons with respect to background activity. Specifically, the clusters of neurons emerged for different tasks in tri-SNN(R) have higher levels of activities than do the rest of the task-irrelevant neurons (nearly resting). By contrast, neuron clusters in tri-SNN(G) have lower levels of activities than do the rest. It is reasonable to believe that these networks are trained to form task-specific neuron clusters characterized by such activity contrasts with their respective backgrounds because the input spike rates encoding specific tasks can be reliably propagated to and decoded by the decision layer (layer 3) while irrelevant information is filtered out. Given this, the reduced (increased) activities of the clustered layer-2 neurons within the transmission line for a specific task in tri-SNN(G) (tri-SNN(R)) would indicate more inhibitory (excitatory) synaptic connections between layer 2 and layer 3. To verify this conjecture, supplementary figure S5 compares the distributions of the weights of the synapses between layer 2 and layer 3 in bi-SNN, tri-SNN(R) and tri-SNN(G) after training on fifty tasks. As expected, the synaptic connections in tri-SNN(G) are inhibitory on average, while those in bi-SNN and tri-SNN(R) are more excitatory.

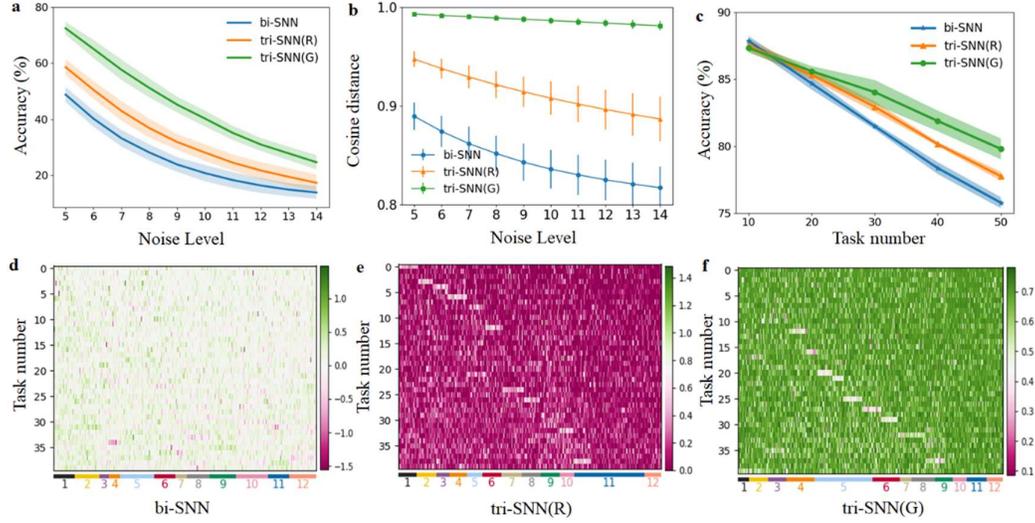

**Figure 4. Comparison of the noise robustness and out-of-distribution representational performance among bi-SNN, tri-SNN(R) and tri-SNN(G). a**. Test accuracies on salt & pepper noise perturbed neuromorphic MNIST data as functions of noise level for bi-SNN, tri-SNN(R) and tri-SNN(G). The curves (shadows) are mean accuracies (standard deviations) over four test trials. **b**. The Cosine distances with respect to the average layer-2 neuron membrane potential between tests on noise corrupted samples and tests on intact ones for these three SNN models. The error bars are standard deviations over four test trials. **c.** Test accuracies averaged over tests on all previously learnt tasks on shuffled MNIST as functions of the number of tasks for these three SNN models. The curves (shadows) are mean accuracies (standard deviations) over four test trials. **d-f.** Learnt neural representations across all tasks for these three SNN models. Neurons are sorted by their cluster membership, indicated by coloured lines at the bottom. K-means clustering method based on task variances is used.

## 2.3. The emulation of $Ca^{2+}$-mediated regulation of synaptic plasticity at dendritic spine in dendritic AAB-transistor

Neural networks can adapt to new environments and various application domains thanks to synaptic plasticity. As the harbors for synapses, dendrites are also the sites where synaptic plasticity occurs. The bi-directional changes of the efficacy of synaptic transmission, namely, long-term potentiation (LTP) and long-term depression (LTD), stabilize networks by avoiding runaway network activity and are thought to be the foundation of memory and learning[54]. The induction of bi-directional synaptic plasticity has generally been believed to depend on $Ca^{2+}$ influx into the postsynaptic dendritic spine: large $Ca^{2+}$ transients have been implicated in the induction of LTP, while smaller $Ca^{2+}$ transients have been associated with LTD[55] (depicted in **figure 5a**). The interconvertible LTP and LTD gives rise to more computational potentials, such as meta-plasticity[56]. In analogy to dendrite, the two interconvertible states of the AAB-transistor under which LTP and LTD are induced, respectively, by positive $V_g$ stimuli are schematically shown in **Figure 5b**.

The experimentally measured AAB transfer curve of the pristine device under cyclic

$V_g$ sweep is shown in **figure 5c**. In the forward (negative-to-positive) $V_g$ sweep, $I_{ds}$ is found to peak in the negative $V_g$ region, around -5 V. In the backward (positive-to-negative) $V_g$ sweep, the $I_{ds}$ peak is found to shift to the positive side of $V_g$, around +0.5 V. What underlies this hysteretic phenomenon is the memristive effect that has been widely exploited for emulating synaptic plasticity[11,57]. The memristive mechanism in our device is illustrated in supplementary figure S6 and explained in supplementary note 5. Because the transfer curve obtained under the forward $V_g$ sweep intersects the $I_{ds}$ axis at a point of lower $I_{ds}$ value than does it obtained under the backward $V_g$ sweep, LTP (LTD) can be inducted in the pristine device by positive (negative) $V_g$ stimuli. To verify, single pulse (+10 V in amplitude, 10 ms in width) and pulse train (1 ms in width and interval) measurements are performed. As shown in **figure 5d**, a single positive pulse results in 0.22 nA increase in the base current (under $V_g$=0 V) and this increase is steady (long-term). Moreover, as shown in **figure 5e**, under a train of pulses $I_{ds}$ increases with the number of pulses, a phenomenon that can be interpreted as the enhanced efficacy of synaptic transmission due to LTP.

Conventionally, LTP-LTD conversion in neuromorphic transistor embodiments is only realizable either in a volatile manner using additional control terminals[58], or by eliciting completely different physical process (i.e., electronic or ionic) for each form of plasticity[59], which is not bio-realistic and adds considerable complexity to the neuromorphic circuits. By contrast, LTP-LTD conversion can be triggered in our dendritic AAB-transistor in a nonvolatile and bio-explainable manner. To elicit LTP-LTD conversion, a sufficiently large conditioning gate bias (CGB) of +40 V is applied for a period of time of the order of tens of seconds. After the CGB has ceased, the hysteretic transfer curve of the device is measured as before. **Figure 5f** shows the transfer curve obtained after 90-s CGB has been applied. A stark contrast to what is shown in figure 5c is observed that the intersection point in the forward $V_g$ sweep is now at a larger value than is in the backward $V_g$ sweep, indicating that LTP (LTD) can be inducted by negative (positive) $V_g$ stimuli. This is in opposite to the situation in the pristine device and can be understood by referring to supplementary figure S6. A single positive pulse results in 0.12 nA decrease in the base current (**figure 5g**). As shown in **figure 5h**, under a train of pulses, $I_{ds}$ decreases with the number of pulses due to LTD. A non-plastic device state can also be achieved by reducing the duration of the positive CGB to ~30 s (supplementary figure S7). The dependence of the difference between the two intersection points in the cyclic $V_g$ sweep on the duration and amplitude of the applied CGB are summarized in supplementary figure S8. Though out of the main scope of this work, a footnote we would like to add is that the demonstrated nonvolatile modulation protocol using CGB is also applicable to the induction of long-term plasticity in recently proposed Gaussian synaptic transistor (has no biological counterpart) for probabilistic neural networks[60], simplifying its device structure and potentially mitigating its power consumption (see supplementary figure S9-11 and discussions in supplementary note 6).

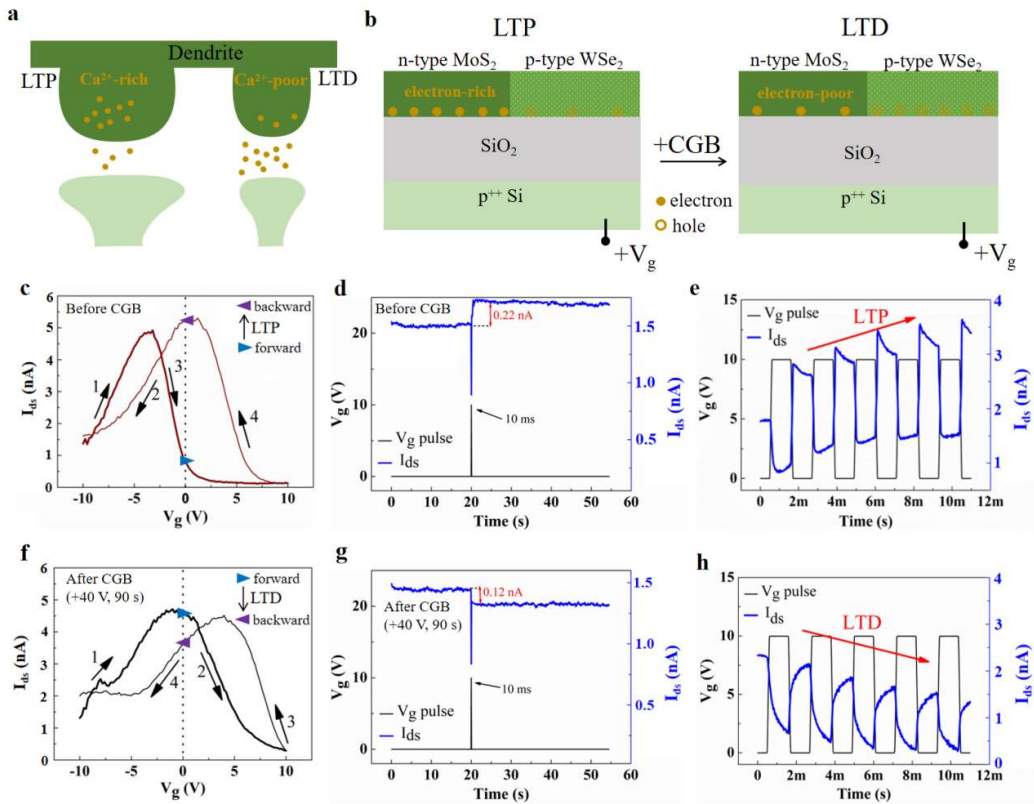

**Figure 5. Reconfigurable hysteretic AAB transfer curve and the emulation of $Ca^{2+}$-mediated synaptic meta-plasticity. a.** Schematic of a biological dendrite as the harbor of synapses and the site where $Ca^{2+}$-mediated bidirectional synaptic plasticity occurs. **b.** Schematic of the two states of the AAB-transistor under which LTP and LTD are inducted, respectively, by positive $V_g$ stimuli. **c.** Hysteretic transfer characteristic of a pristine dendritic AAB-transistor without CGB and its corresponding LTP character under **d.** single pulse measurement and **e.** pulse train measurement. **f.** Right shift of the hysteretic AAB transfer curve after CGB of + 40 V for 90 s has been applied and its corresponding LTD character under **g.** single pulse measurement and **h.** pulse train measurement.

As an emulator of dendritic functional phenomenology, the workings of the AAB-transistor are surprisingly similar to those of its biological counterpart at the subcellular level, in contrast to previous devices[30-34]. Recall that the LTP-LTD conversion takes place after the CGB which induces substantial carrier exchange through the channel/dielectric interface (supplementary figure S6). This is in analogy to the $Ca^{2+}$ exchange through the dendritic membrane. Thus, it is reasonable to interpret carriers in our device biologically as the $Ca^{2+}$. Previously, analogy has been drawn between $Ca^{2+}$ and mobile ions in two-terminal synaptic devices in interpreting short-term synaptic plasticity[61]. Unfortunately, the most frequently emulated long-term synaptic plasticity has lacked biological explainability at the subcellular level. Our dendritic AAB-transistor fills this gap. Given this, when re-considering the emulation of dCaAP, it is not hard to find that the emulated dCaAP also has very similar physical origin to the biological one, that is, free carrier generation in the channel in our device in analogy to $Ca^{2+}$ influx into the membrane in a biological dendrite.

## 3. Conclusion

To conclude, we have demonstrated a dendritic AAB-transistor with functional versatility and biological explainability, filling a gap in neuromorphic electronics. This dendritic transistor is based on heterojunction channel, giving rise to unique anti-ambipolar transfer characteristic that can be used to emulate the recently discovered non-monotonic dCaAP, enabling the implementation of complex inference functions in a single device, such as the logic XOR function that has traditionally been deemed solvable only by multi-layer neural networks. This dendritic AAB-transistor, along with the existing neuro-transistor and synaptic transistor, has the potential to be used in fully transistor-implemented hardware accelerators for bio-realistic SNNs. By regarding SNN as a synapse-dendrite-neuron trinity, we demonstrate that tri-SNN(G) with nonconventional but bio-plausible non-monotonic Gaussian-like dAF surpasses bi-SNN and tri-SNN(R) with monotonic ReLU dAF in terms of representational capability, noise robustness and continual learning capability. Visualization of the inner workings of the neural network models and quantitative interpretability analyses reveal that non-monotonic dendritic action potential significantly affects spike activity dynamics, which in turn contributes to respectable performance boost. In addition to inference functions, learning functions (i.e., plasticity) can also be emulated in this dendritic AAB-transistor device based on its hysteretic AAB transfer characteristic due to the memristive effects. The learning functions also cover a full range, from long-term potentiation to long-term depression, mimicking meta-plasticity. A parallel can be drawn between the electronic origins of these mimicked dendritic functions and the biochemical origins of their biological counterparts, indicating high bio-fidelity of our device. This work represents an exciting progress in the fields of neuroscience-inspired machine intelligence and post-Moore nanotechnology, and may stimulate interest in ANN reinvention at its most fine-grained level toward human-level intelligence.

## 4. Methods

*MoS$_2$ and WSe$_2$ thin-film deposition:* MoS$_2$ and WSe$_2$ films are deposited by magnetron sputtering (AJA, USA) via radio-frequency (RF) mode on highly p-type doped Si substrates (as the back-gates) with 100 nm thermally grown SiO$_2$. Both targets are stoichiometric. Before sputtering, the substrates are heated to 80 °C. Then, all deposition processes are conducted at room temperature. RF powers of 15 W and 30 W are used for MoS$_2$ and WSe$_2$ depositions, respectively, under an Ar (99.999%) flow rate of 10 sccm. The base and working pressures of the sputtering chamber are ~ $1.0 \times 10^{-7}$ Torr and ~ $3.0 \times 10^{-3}$ Torr, respectively. The samples are in-situ annealed under 220 °C for 1 hour after deposition.

*Dendritic anti-ambiplor (AAB)-transistor fabrication:* The MoS$_2$ film is first deposited and then the heterojunction area is defined by ultra-violet lithography, followed by the deposition of WSe$_2$. After that, the heterojunction is annealed under 800 °C for 2 hours in the Ar atmosphere. The source and drain electrodes of 35-nm Pt are then patterned by the conventional photolithography process and deposited by DC sputtering (30 W, $3.0 \times 10^{-3}$ Torr), followed by the lift-off steps. The channel lengths of the two control homojunction-channel devices are 15 μm, which are the same as that of the heterojunction-channel transistor.

*Materials characterization:* AFM analyses are conducted by an atomic force microscope (DIMENSION ICON, BRUKER, USA) under the tapping mode. Raman spectra are obtained on a single-gating micro-Raman spectrometer (Horiba-JY T64000) excited with 532 nm laser. XPS measurements are performed in a PHI Quantera II system.

*Electrical measurements:* The quasi-DC voltage sweeps are performed by the Keysight B1500A semiconductor analysis system equipped with high-resolution source and measurement units with a specified resolution of 1 fA. The Keysight B1530A waveform generator/fast measurement unit is used to perform the pulse measurements.

*SNN model parameterization, training and tests:* To simplify implementation, each neuron is modelled to have only one dendritic branch ($D^i$ =1). For pattern classification tasks on Fashion-MNIST[44], the adopted network structure is [784-FC512-FC10] (FC: fully-connected layer). The time window $T$ is set to 10 and the threshold $V_{th}$ is set to 0.75. The minimum time step $\Delta t$ is set to 1. The models are optimized using the adaptive moment estimation (Adam) optimizer[62] with an initial learning rate of 0.0005 and training with batch size of 100. Gradient substitution method is adopted for the backpropagation of mean square error (MSE) and rectangular function is used to approximate the derivative of spike activity[45] with the rectangular length of 0.5. All simulations are performed by PyTorch on 4 RTX 2080Ti GPUs.

For noise-robustness tests, the adopted network structure is [2312-FC512-FC10] and the models are optimized by Adam optimizer. The time window $T$ is set to 10 and the threshold $V_{th}$ is set to 0.75. The three SNN models are pre-trained by minimizing the MSEs on 50000 intact Neuromorphic-MNIST[51] training samples for 100 epochs and then tested on 1000 Neuromorphic-MNIST test samples contaminated by salt & pepper noise with varying intensities measured by the percentage of the perturbed pixels.

For continual learning tests, the network structure is [784-FC1024-FC10]. The time window $T$ is set to 5 and the threshold $V_{th}$ is set to 0.5. Tasks are to classify handwritten digits 0-9. The training sample images are distorted by randomly permuting the pixels. Each consecutive task adopts a different randomization protocol, on which the models are trained with Adam by minimizing the MSEs in ten epochs. In training on each task, 3% randomly selected synaptic connections are active. Afterwards, the abilities of the networks to perform the previously learnt tasks are evaluated on the corresponding test samples, with the 3% synaptic paths selected in training the corresponding tasks reused.

**Supporting Information**
Supporting Information is available from the Online Library or from the author.

**Acknowledgments**

Y.Y. and M.X. contributed equally to this work. H.L. conceived the idea and supervised the project. Y.Y. and P.L. performed the device fabrication and measurements. M.X., J.P., G.L. and S.W. conducted the simulations. Y.Y., M.X. and H.L. wrote this manuscript. This research was supported by National Natural Science Foundation (grant nos. 61974082, 61704096, 61836004), National Key R&D Program of China (2021ZD0200300, 2018YFE0200200), Youth Elite Scientist Sponsorship (YESS) Program of China Association for Science and Technology (CAST) (no. 2019QNRC001), Tsinghua-IDG/McGovern Brain-X program, Beijing science and technology program (grant nos. Z181100001518006 and Z191100007519009), Suzhou-Tsinghua innovation leading program 2016SZ0102, CETC Haikang Group-Brain Inspired Computing Joint Research Center.